%% file: main.tex
\pdfoutput=1  
\documentclass[pdflatex,sn-basic]{sn-jnl}

\usepackage{graphicx}
\usepackage{amsmath,amssymb}
\usepackage{booktabs}
\usepackage{tabularx}

\newcolumntype{L}[1]{>{\raggedright\arraybackslash\hsize=#1\hsize}X}

\graphicspath{{figures/}}

\allowdisplaybreaks
\begin{document}

\title[]{Perturbational Validity for Foundation Models of Brain Dynamics: A Controlled Proof-of-Principle Simulation}

\author[1,2,3]{\fnm{Jos\'e C.} \sur{Garc\'ia Alanis}}
\author[1,3]{\fnm{Sarah} \sur{Alizadeh}}
\author[1]{\fnm{Marco} \sur{Rothermel}}
\author[1]{\fnm{Bita} \sur{Shariatpanahi}}
\author[2]{\fnm{Mina} \sur{Kheirkhah}}
\author[2]{\fnm{Stefan G.} \sur{Hofmann}}
\author[4]{\fnm{Tim} \sur{Hahn}}
\author*[1,3]{\fnm{Hamidreza} \sur{Jamalabadi}}\email{hamidreza.jamalabadi@uni-marburg.de}

\affil[1]{\orgdiv{Department of Psychiatry and Psychotherapy}, \orgname{Philipps-Universit\"at Marburg}, \country{Germany}}
\affil[2]{\orgdiv{Department of Psychology}, \orgname{Philipps-Universit\"at Marburg}, \country{Germany}}
\affil[3]{\orgdiv{Center for Mind, Brain, and Behavior (CMBB)}, \orgname{Philipps-Universit\"at Marburg}, \country{Germany}}
\affil[4]{\orgdiv{Institute for Translational Psychiatry}, \orgname{University of M\"unster}, \country{Germany}}

\makeatletter
\g@addto@macro\auaddress{%
  \addvspace{3pt}%
  {ORCID: Jos\'e~C.~Garc\'ia~Alanis,
   \href{https://orcid.org/0000-0003-3459-222X}{0000-0003-3459-222X};
   Hamidreza~Jamalabadi,
   \href{https://orcid.org/0000-0003-2485-2374}{0000-0003-2485-2374}.\par}%
}
\makeatother

\abstract{Foundation models for human brain recordings are usually evaluated by signal reconstruction, future-state prediction, and transfer to downstream tasks. However, these benchmarks do not establish whether a transferred model remains valid when the system is actively perturbed. We define perturbational validity as the preservation, after limited system-specific adaptation, of the conditional distribution of future trajectories given the current state and a controlled input, and we evaluate it at three levels: time-series accuracy, dynamical-structure similarity, and responses to perturbations excluded from calibration. We demonstrate the framework in an oracle-drift simulation of stochastic bistable systems. Two otherwise identical multilayer perceptrons were trained on drift evaluations from passive or input-driven trajectories. Next, for each held-out system the shared weights were frozen and only a three-dimensional embedding was adapted, with a correctly specified cubic model fitted from scratch as comparator. With two to five system-specific evaluations, perturbational pretraining yielded lower errors in recovering controlled flow, landscape geometry, finite-run occupancy, response distributions, and dose-transition curves. The advantage was reproduced across five independent runs, persisted under full-network adaptation of the passive model, and was attributable to input excitation rather than transition-state coverage: excitation alone lowered controlled-flow error 1.94-fold relative to coverage alone, in five of five runs. The cubic model became competitive as calibration grew, showing that the benefit is specific to few-shot transfer. This controlled demonstration does not test recovery of dynamics from noisy or partially observed brain recordings. It shows why passive prediction should be complemented by prospective evaluation under controlled inputs.
}


\maketitle

\input{sections/01_introduction}
\input{sections/02_framework}
\input{sections/03_simulation}
\input{sections/04_discussion}
\input{sections/05_conclusion}

\input{sections/06_declarations}

\bibliography{references}

\ifdefined\nosupplement\else
  \clearpage
  \input{supplementary/supplement}
\fi

\end{document}

%% file: sections/01_introduction.tex
\section{From observation to intervention}
\label{sec:introduction}

The central promise of a brain foundation model is transfer: structure learned from large population datasets should support adaptation to new participants, tasks, and recording settings when target data are limited \citep{Bommasani2021-at}. Recent observationally pretrained models illustrate this potential across modalities. For electroencephalography (EEG), REVE uses masked autoencoding on more than 60,000 hours of recordings and coordinate-based positional encoding to accommodate different electrode configurations \citep{elouahidi2025reve}. For functional magnetic resonance imaging (fMRI), BrainLM uses masked prediction to support future-state forecasting and clinical prediction \citep{Ortega-Caro2024-uy}, whereas Brain-JEPA combines joint-embedding prediction with functional-gradient positioning and spatiotemporal masking \citep{dong2024brainjepa}. Together, these models show that large observational datasets can yield representations that transfer across participants and support multiple downstream tasks, that is, prediction problems distinct from the pretraining objective.

Most evaluations of human EEG and fMRI foundation models, however, emphasize masked-signal reconstruction, short-horizon forecasting, linear probing, classification, or regression \citep{elouahidi2025reve,Ortega-Caro2024-uy,dong2024brainjepa}. These benchmarks establish representational utility within an observational or task-specific distribution, but not whether an adapted model predicts how dynamics change under controlled input. In animal systems neuroscience, by contrast, \citet{wang2025foundation} showed that a foundation model of mouse visual cortex, pretrained on stimulus-driven responses pooled across animals, transferred to new mice from little data and predicted responses to stimulus domains absent from training. Closed-loop experiments with such models indicate that a model fitted under one input regime can guide the design of an intervention, but that its predictions for new inputs must be verified prospectively, because such models are not guaranteed to generalise beyond their training statistics \citep{walker2019inception}. This matters for human neuromodulation, where stimulation changes multiregional brain activity in a parameter-dependent way that a model must predict for each individual \citep{yang2021modelling}. None of these studies, however, asks whether a population model, after limited adaptation to a new individual, preserves the controlled dynamics that determine how an intervention moves the system.

The question cannot be settled from passive data alone, for a reason that classical system identification makes precise: recovering an input-response relationship requires sufficiently informative inputs \citep{Ljung1999-bv,willems2005note}. Passive recordings constrain autonomous dynamics mainly along spontaneously visited trajectories, leaving rarely visited states and transition regions weakly determined. More fundamentally, when the passive condition is $u(t)=0$, parameters that affect the dynamics only through terms multiplied by $u$ are absent from the passive data-generating process. Systems with different input gains can therefore generate identical passive trajectories while implying different reachable states, transition thresholds, and responses to intervention. Observational evaluation cannot detect this difference, because the behaviour that distinguishes such systems never appears in the data it scores. What a transferred model must additionally preserve is the input-dependent law itself, a property we call \emph{perturbational validity}.

Stated precisely, following limited adaptation to a new participant, a perturbationally valid model preserves the conditional distribution of future trajectories given the current state and a controlled input. This criterion does not imply complete biological identification. Hidden states, measurement noise, partial observation, and non-stationarity may prevent recovery of the underlying mechanism \citep{Durstewitz2023-ta}. Perturbational validity is instead a prospective test: does the transferred model remain predictive when the system is driven by inputs excluded from participant-specific calibration?

\begin{figure}[!p]
\centering
\includegraphics[width=\textwidth,height=0.82\textheight,keepaspectratio]{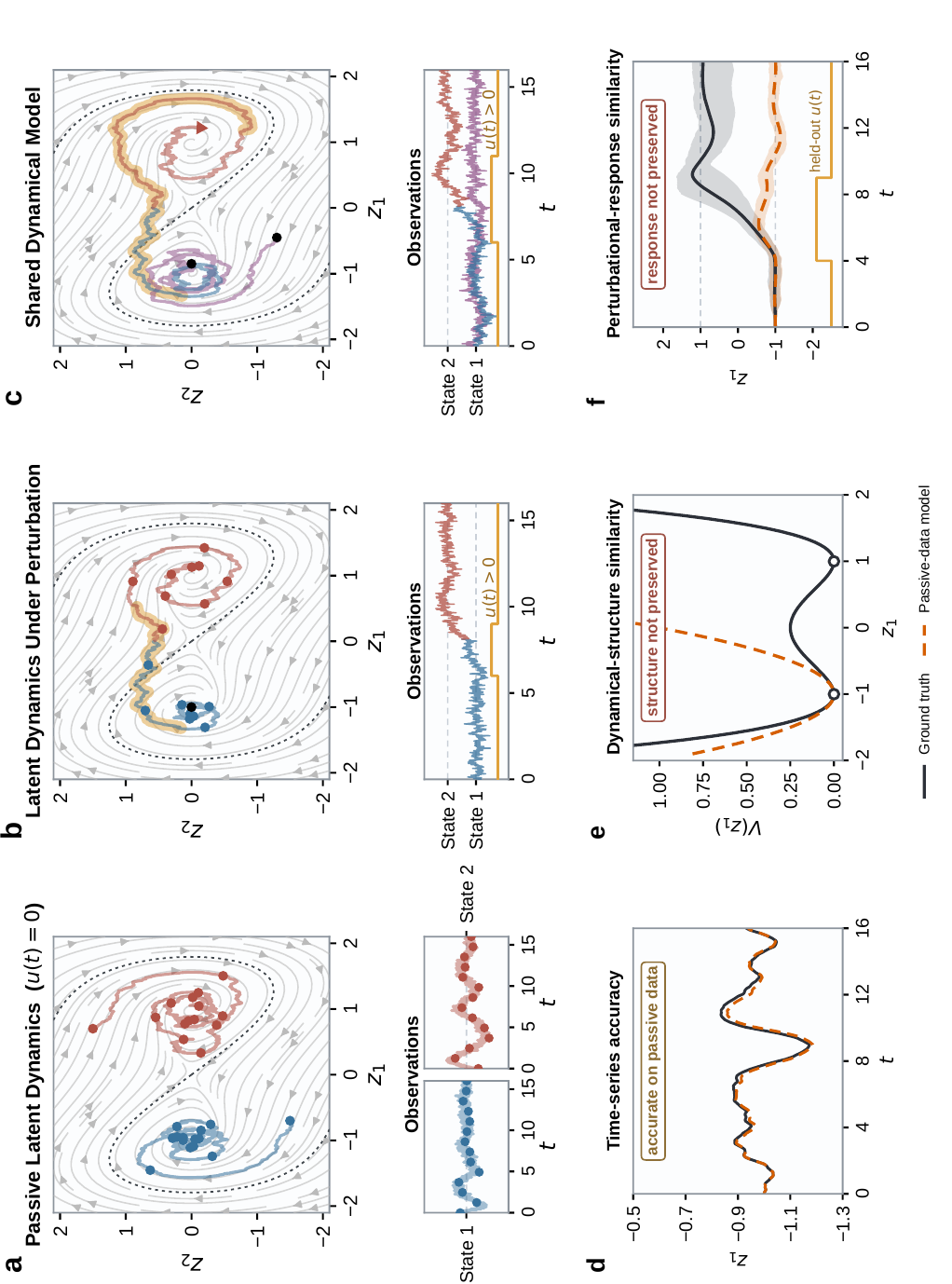}
\caption{\textbf{Perturbational validity for transferred dynamical models.}
\textbf{a}, Passive observations concentrate around frequently occupied states and do not identify input coupling when the input is absent.
\textbf{b}, A controlled input drives the latent state across a basin boundary, revealing behaviour absent from the passive record.
\textbf{c}, Systems with the same unforced dynamics but different input gains produce identical passive behaviour yet respond differently to the same perturbation.
\textbf{d-f}, A model fitted locally around the observed attractor reproduces the passive trajectory (\textbf{d}) but fails to preserve the second attractor and transition barrier (\textbf{e}) and consequently predicts the wrong response to a held-out input (\textbf{f}). Perturbational validity is therefore evaluated through complementary evidence from time-series accuracy, dynamical-structure similarity, and perturbational-response similarity}
\label{fig:concept}
\end{figure}

Figure~\ref{fig:concept} illustrates why this criterion is needed. Passive recordings primarily sample frequently occupied states and contain no information about input coupling when $u=0$ (Fig.~\ref{fig:concept}a). Perturbations can reveal transitions absent from the passive record (Fig.~\ref{fig:concept}b), while systems with identical unforced dynamics may respond differently because their input gains differ (Fig.~\ref{fig:concept}c). The counterexample in Fig.~\ref{fig:concept}d-f further shows that reproducing an observed passive trajectory does not guarantee preservation of a second attractor, its transition barrier, or the response to a held-out input.

These are the properties that matter when a model is meant to guide an intervention, because an intervention does not merely reveal the system; it changes its trajectory. In dynamical accounts of brain and behaviour, intervention outcomes depend on attractor location and stability, transition geometry, and the coupling between external inputs and internal states \citep{Scheffer2024-mq,Gu2015-qx,Tang2018-uc}. These ideas have motivated proposals to treat psychiatric interventions as control inputs and to use targeted N-of-1 perturbations to construct individualized surrogate models for counterfactual prediction and treatment selection \citep{Hofmann2025-mm,kheirkhah2026reengineering}. Such applications require more than accurate passive forecasts: models with similar short-horizon prediction errors can differ in attractor geometry and long-run statistics \citep{Hemmer2025-xa}, and therefore in the treatment responses they predict.

We therefore evaluate perturbational validity at three complementary levels. Time-series accuracy assesses predictions over a defined horizon. Dynamical-structure similarity assesses the controlled vector field and associated generative properties, including attractors, transition geometry, and long-horizon occupancy. Perturbational-response similarity assesses response distributions, recovery dynamics, transition probabilities, and dose-response relationships under inputs excluded from participant-specific calibration. No individual level is sufficient, and no finite collection of metrics can establish equivalence over every state and input. Together, however, these levels provide stronger evidence of intervention-relevant transfer than passive prediction alone.

We formalise this framework using stochastic controlled latent dynamics and, where a gradient representation exists, an effective energy landscape. We then test it in a deliberately minimal family of stochastic bistable systems. Two otherwise identical population models are trained on analytical oracle drift targets sampled from either passive or perturbational trajectories across 60 training systems. For each of 20 held-out systems, the shared network parameters are frozen and only a three-dimensional system-specific embedding is adapted. A correctly specified cubic model fitted from scratch provides a non-pretrained comparator.

Perturbational pretraining provides its clearest advantage when only two to five system-specific drift evaluations are available, improving recovery of the controlled vector field, effective energy landscape, stochastic response distribution, and dose-transition relationship relative to passive pretraining. The correctly specified from-scratch model becomes competitive as calibration increases, showing that the observed benefit is specific to few-shot transfer rather than a superior asymptotic solution. Supplementary Tables~S1 and S2 provide the complete simulation and optimization specifications. Supplementary Fig.~S1 shows that the few-shot pattern is reproduced across five independent end-to-end runs, persists when the passive model is allowed full-network adaptation, is driven more strongly by input excitation than by transition-state coverage alone, and extends to one pulse amplitude just outside population-pretraining support. The latter result supports local extrapolation only.

The study therefore introduces an evaluation framework rather than a new foundation-model architecture. Its central proposal is that models intended to support experimentation or intervention should be evaluated not only by observational transfer, but also by whether they preserve the consequences of actively driving the system.

%% file: sections/02_framework.tex
\section{A perturbational framework for transfer}
\label{sec:framework}

\subsection{Problem statement}

A model intended for intervention must solve a stronger transfer problem than a model intended only for reconstruction or short-term forecasting. Let $\mathbf{z}_s(t)$ denote the latent state of participant $s$, $\mathbf{u}(t)$ an experimentally controlled input, and $\mathbf{y}_s(t)$ the measured signal. Population pretraining provides shared structure, whereas deployment requires adaptation to a new participant from a small calibration set. In the general case, calibration requires observed state transitions or trajectory segments, which we write in discrete form as
\[
\mathcal{D}^{\mathrm{cal}}_s
=
\left\{
\left(
\mathbf{y}_s(t_i),
\mathbf{u}(t_i),
\mathbf{y}_s(t_i+\Delta t)
\right)
\right\}_{i=1}^{n_{\mathrm{cal}}}.
\]
An adaptation procedure $A_{\theta}$ uses these data to obtain a participant representation,
\[
\widehat{\mathbf{e}}_s
=
A_{\theta}
\left(
\mathcal{D}^{\mathrm{cal}}_s
\right).
\]

In the proof-of-principle simulation below, the state is observed directly and calibration uses analytical drift evaluations rather than finite-difference transitions:
\[
\mathcal{D}^{\mathrm{cal}}_s
=
\left\{
\left(
x_i,
u_i,
f_s(x_i,u_i)
\right)
\right\}_{i=1}^{n_{\mathrm{cal}}}.
\]
This oracle-drift setting deliberately removes state-estimation and derivative-estimation error so that the experiment isolates transfer of the controlled dynamics.

The transfer objective is to recover participant-specific controlled dynamics well enough to predict the state distribution at a specified horizon under inputs that were not used for participant-specific calibration:
\begin{equation}
p\left(
\mathbf{z}_s(t+\tau)
\mid
\mathbf{z}_s(t),
\mathbf{u}_{t:t+\tau},
\widehat{\mathbf{e}}_s
\right).
\label{eq:transfer-target}
\end{equation}
Here, ``held out'' means excluded from participant-specific calibration; an input may still have been represented during population pretraining. Perturbational validity is therefore relative to the calibration budget, prediction horizon, and state-input domain over which Eq.~\ref{eq:transfer-target} is evaluated.

The central difficulty is identifiability. Passive observations constrain the dynamics mainly near frequently occupied states, leaving transition regions and behaviour outside the observational support weakly determined. Input coupling presents a stronger problem. Throughout this work, the passive condition is $\mathbf{u}\equiv\mathbf{0}$, and input-dependent parameters enter only through terms multiplied by $\mathbf{u}$. The within-participant passive likelihood therefore contains no information about these parameters: two systems that differ only in input gain can generate exactly the same passive trajectories \citep{Ljung1999-bv,Villaverde2019-qf,schoukens2019nonlinear}. Population correlations could provide indirect prior information about input response, but the passive observations themselves do not identify it.

Perturbational population data address this problem in two ways: they expose the shared model to how inputs alter the dynamics and increase coverage of states that are rarely visited passively. The resulting population model can learn reusable input-response structure, while limited participant-specific calibration locates a new participant within that family. Passive pretraining alone cannot identify this shared input dependence. This does not imply that input response could never be learned from sufficiently rich participant-specific perturbation data; the claim concerns efficient transfer when participant-specific calibration is limited.

\subsection{Controlled latent dynamics}

We represent the population model as a stochastic controlled system,
\begin{equation}
d\mathbf{z}_t
=
\mathbf{f}_{\theta}
\left(
\mathbf{z}_t,
\mathbf{u}_t,
\mathbf{e}_s
\right)dt
+
\boldsymbol{\Sigma}_{\theta}
\left(
\mathbf{z}_t,
\mathbf{u}_t,
\mathbf{e}_s
\right)d\mathbf{W}_t,
\label{eq:latent-dynamics}
\end{equation}
with observation model
\begin{equation}
\mathbf{y}_s(t)
=
\mathbf{h}_{\theta}
\left(
\mathbf{z}_s(t)
\right)
+
\boldsymbol{\epsilon}_s(t).
\label{eq:observation}
\end{equation}
The shared parameters $\theta$ encode population-level structure. In the present study, adaptation to a new participant is restricted to the low-dimensional representation $\mathbf{e}_s$, while all shared parameters remain frozen. More general adaptation procedures or posterior distributions over participant-specific parameters are possible but are not examined here.

The intervention-relevant object is the controlled transition law in Eq.~\ref{eq:transfer-target}. Locally, this law is characterised by the controlled drift $\mathbf{f}(\mathbf{z},\mathbf{u})$ together with the diffusion. For the stochastic system in Eq.~\ref{eq:latent-dynamics}, these components define the infinitesimal generator
\begin{equation}
\mathcal{L}^{\mathbf{u}}\phi(\mathbf{z})
=
\mathbf{f}(\mathbf{z},\mathbf{u})^{\top}
\nabla\phi(\mathbf{z})
+
\frac{1}{2}
\operatorname{tr}
\left[
\mathbf{D}(\mathbf{z},\mathbf{u})
\nabla^2\phi(\mathbf{z})
\right],
\qquad
\mathbf{D}
=
\boldsymbol{\Sigma}
\boldsymbol{\Sigma}^{\top}.
\label{eq:generator}
\end{equation}
For a time-varying input, $\mathcal{L}^{\mathbf{u}}$ is understood pointwise along the specified input sequence. Passive data constrain this generator only along spontaneously visited trajectories under $\mathbf{u}=\mathbf{0}$. Intervention requires it to generalise to state-input combinations that are relevant to control but absent from participant-specific calibration \citep{Pavliotis2014-id}.

The proof-of-principle simulation considers a simpler setting than Eq.~\ref{eq:latent-dynamics}: the state is observed directly, the observation map is the identity, measurement noise is absent, and the diffusion amplitude is fixed and known. The simulation therefore evaluates transfer of the controlled drift rather than simultaneous recovery of latent state, observation map, drift, and diffusion.

\subsection{Energy-landscape perspective}

For gradient systems such as the proof-of-principle model, the controlled dynamics can be organised through an effective energy function,
\begin{equation}
\mathcal{H}
\left(
\mathbf{z},
\mathbf{u};
\mathbf{e}_s
\right)
=
V
\left(
\mathbf{z};
\mathbf{e}_s
\right)
-
\mathbf{u}^{\top}
\mathbf{G}
\left(
\mathbf{z};
\mathbf{e}_s
\right),
\label{eq:effective-hamiltonian}
\end{equation}
with
\begin{equation}
\dot{\mathbf{z}}
=
-\mathbf{M}(\mathbf{z})
\nabla_{\mathbf{z}}
\mathcal{H}
\left(
\mathbf{z},
\mathbf{u};
\mathbf{e}_s
\right).
\label{eq:gradient-flow}
\end{equation}
Here, $V$ describes the unforced landscape, $\mathbf{G}$ contains potential-coupling functions for the inputs, and $\mathbf{M}$ is a symmetric positive mobility matrix. For a time-varying input, $\mathcal{H}(\mathbf{z},\mathbf{u}_t)$ is an instantaneous effective energy that changes as the input changes. It is not a Hamiltonian for conservative mechanics.

The one-dimensional simulation is exactly of this form. There, $\mathbf{z}=x$, $\mathbf{M}=1$, and $G_s(x)=b_sx$, giving
\[
\mathcal{H}_s(x,u)
=
V_s(x)-b_sux,
\qquad
f_s(x,u)
=
-\frac{\partial\mathcal{H}_s(x,u)}{\partial x}.
\]
The input therefore tilts the bistable potential, changing the effective barrier and the probability of transition between basins.

This representation separates three intervention-relevant quantities: the locations and local curvature of stable states, the barriers separating them, and the way inputs deform those barriers. Passive trajectories can constrain the occupied minima while leaving transition geometry and input coupling underdetermined. Perturbations can provide both additional state-space coverage and the input excitation needed to constrain these quantities.

A unique scalar potential need not exist in higher-dimensional or non-gradient systems, which are therefore not described by Eq.~\ref{eq:gradient-flow}: their drift generally contains a component that no single scalar $\mathcal{H}$ generates. In such cases, the relevant validation targets are the controlled vector field, stochastic generator, quasipotential, committor function, or transition operator \citep{Freidlin2012-wc,nolting2016balls,E2010-xz}. Perturbational validity is defined on the controlled transition law of Eq.~\ref{eq:transfer-target} and therefore does not require an energy landscape; Eqs.~\ref{eq:effective-hamiltonian} and~\ref{eq:gradient-flow} are a transparent special case, satisfied exactly by the simulated system.

\subsection{Evaluation criteria}

The complete controlled transition law cannot be evaluated exhaustively from a finite test set. We therefore examine three complementary groups of observable consequences---realised trajectories, dynamical structure, and responses to held-out perturbations---summarised in Table~\ref{tab:transfer-metrics}.

\begin{sidewaystable}[p]
\caption{\textbf{Three complementary layers for evaluating perturbational transfer.}
Time-series metrics assess realised predictions, structural metrics assess the recovered generative dynamics, and perturbational metrics assess the consequences of controlled inputs. Together, the three layers provide complementary evidence of intervention-relevant transfer.}
\label{tab:transfer-metrics}
\setlength{\tabcolsep}{5pt}
\renewcommand{\arraystretch}{1.2}

\begin{tabularx}{\textheight}{
L{0.75}
L{0.95}
L{1.05}
L{1.45}
L{0.80}}
\toprule
\textbf{Evaluation layer} &
\textbf{Core question} &
\textbf{Metrics reported here} &
\textbf{Other suitable metrics} &
\textbf{Evidence of transfer} \\
\midrule

\textbf{1. Time-series accuracy} &
Does the model forecast state evolution over a defined horizon? &
Deterministic passive-rollout RMSE against a stochastic reference trajectory &
One- or multi-step error; MAE; $R^2$; correlation; probabilistic scores; spectral or autocorrelation error &
Accurate forecasts over prespecified horizons \\

\addlinespace

\textbf{2. Dynamical-structure similarity} &
Has the model recovered the law and geometry that generate possible trajectories? &
Controlled-flow RMSE; finite-run empirical-occupancy Jensen-Shannon divergence; attractor-location error; illustrative landscape and barrier recovery &
Generator discrepancy; Wasserstein or Jensen-Shannon divergence; fixed points; Jacobian spectra; saddles; quasipotential; transition operators; mean first-passage times &
Correct stability, occupancy, and transition geometry \\

\addlinespace

\textbf{3. Perturbational-response similarity} &
Does an input held out from participant calibration produce the correct change in the system? &
Response-distribution JS; dose-transition RMSE; illustrative transition outcome &
Impulse-response error; recovery time; peak response; committor probabilities; intervention thresholds; control regret &
Correct response distributions and intervention outcomes \\

\bottomrule
\end{tabularx}
\end{sidewaystable}

The three layers answer distinct but connected questions. Time-series accuracy asks whether state evolution is forecast over a prespecified horizon. In the simulation below, this is a deterministic drift rollout compared with one stochastic reference trajectory and therefore has an irreducible process-noise floor. Dynamical-structure similarity asks whether the controlled law and the geometry generating possible paths are recovered. Perturbational-response similarity asks whether that law produces the correct prospective consequences under controlled inputs. Thus, the second layer assesses the recovered mechanism, whereas the third assesses its realised intervention outcomes. A model intended for experimentation or control should therefore not be judged from passive rollout accuracy alone.

%% file: sections/03_simulation.tex
\section{Proof-of-principle simulation}
\label{sec:simulation}

\subsection{Generative system and model classes}

We constructed a deliberately minimal family of stochastic one-dimensional bistable systems. Each synthetic system represents a participant-specific member of a shared population family. Its state evolved according to
\begin{equation}
dx_t
=
\left[
-\left(
a_s x_t^3-c_s x_t-d_s
\right)
+
b_s u_t
\right]dt
+
\sigma_s dW_t,
\label{eq:double-well}
\end{equation}
where $a_s$ controls the quartic confinement, $c_s$ the depth and separation of the wells, $d_s$ their asymmetry, $b_s$ the input gain, and $\sigma_s$ the diffusion amplitude. The scalar state $x_t$ is the latent state $\mathbf{z}_t$ of Section~\ref{sec:framework} reduced to one dimension. We write it as $x$ throughout this section to distinguish the simulated system from the general formulation.

The input gain $b_s$ enters only through the product $b_su_t$. For two systems that differ only in $b_s$, the conditional distribution of passive trajectories under $u_t\equiv0$ is therefore identical, and $b_s$ is not identifiable from within-system passive observations alone without additional population-level assumptions or informative input variation (Section~\ref{sec:framework}).

We observe the state directly, so the observation model of Eq.~\ref{eq:observation} is the identity and $\mathbf{y}_t=x_t$. This deliberate simplification removes measurement error and identification of the observation map $h_{\theta}$ from the comparison. Differences between pretraining regimes therefore reflect what they recover about the dynamics rather than the readout. Recovering latent states from indirect measurements is a separate problem accommodated by the general framework but not tested here.

In the absence of input, the drift is generated by
\begin{equation}
V_s(x)
=
\frac{a_s}{4}x^4
-
\frac{c_s}{2}x^2
-
d_s x,
\qquad
f_s(x,0)
=
-\frac{dV_s}{dx}.
\label{eq:potential}
\end{equation}
The input adds the linear tilt $-b_sux$ to the effective energy,
\begin{equation}
\mathcal{H}_s(x,u)
=
V_s(x)-b_sux,
\qquad
f_s(x,u)
=
-\frac{\partial\mathcal{H}_s(x,u)}{\partial x}.
\label{eq:controlled-potential}
\end{equation}
This system isolates the distinction introduced above. Passive observations primarily constrain the unforced landscape near the occupied wells, whereas perturbations additionally reveal the transition region and the participant-specific input coupling.

The population contained 60 training systems and 20 held-out systems. Parameters varied across systems as
\[
a_s=1,
\qquad
c_s\in[0.75,1.30],
\qquad
d_s\in[-0.28,0.28],
\qquad
b_s\in[0.60,1.45],
\qquad
\sigma_s=0.14.
\]
Draws were accepted only when $27a_sd_s^2<4c_s^3$, the condition under which
$f_s(x,0)$ has three roots and $V_s$ therefore has two wells separated by a
barrier. The declared ranges admit a corner of large $|d_s|$ and small $c_s$ in
which one well disappears; rejecting those draws makes every sampled system
bistable without narrowing the ranges, so basin-transition metrics always refer
to a switch between two states.

Dynamics were simulated by Euler-Maruyama integration with $dt=0.02$. Each training system contributed 1,500 time steps after burn-in. Passive pretraining used $u_t=0$. Perturbational pretraining used random pulses with varied sign, amplitude, onset, and duration, superimposed on low-amplitude input noise and generated from both initial basins. The passive and perturbational datasets contained the same number and length of trajectories and were generated from the same population family; the defining difference was the presence or absence of controlled excitation. Supplementary Table~S1 provides the complete generative, sampling, calibration, and evaluation specification.

To isolate transfer of the controlled drift from drift-estimation error, both population pretraining and participant calibration used the analytical drift $f_s(x_t,u_t)$ as the supervised target rather than a finite-difference estimate from the simulated trajectory. The experiment is therefore an oracle-drift proof of principle and does not test drift recovery from noisy state increments.

The two population models used the same architecture. A multilayer perceptron with two tanh hidden layers of 24 units predicted the instantaneous drift from the current state, current input, and a three-dimensional participant embedding,
\begin{equation}
\widehat{f}_{\theta}
\left(
x,u,\mathbf{e}_s
\right).
\end{equation}
One model was pretrained on passive trajectories and the other on perturbational trajectories. For each held-out system, all shared network weights were frozen and only $\mathbf{e}_s$ was estimated from oracle drift evaluations. Thus, the primary comparison tests low-dimensional embedding adaptation rather than full fine-tuning of a passively pretrained model. Supplementary analyses additionally allowed all weights of the passive model to adapt during calibration.

The embedding dimension equals the number of participant-varying drift parameters, $c_s$, $d_s$, and $b_s$, making this a deliberately favourable low-dimensional transfer setting. The embedding coordinates were not constrained to recover these parameters one-to-one. Full architecture, optimization, adaptation, and regularization details are given in Supplementary Table~S2.

Calibration sizes were
\[
n_{\mathrm{cal}}
\in
\left\{
2,3,5,8,12,20,35,60,100
\right\}.
\]
For each size, calibration evaluations were sampled independently rather than as nested subsets. Approximately half were selected, where possible, from perturbed periods operationally defined by $|u|>0.15$; the remainder came from periods with $|u|\leq0.15$. A regularised cubic drift model fitted only to the held-out system served as the from-scratch baseline. This baseline had the correct polynomial form, including a linear input term, and was fitted in physical coordinates so that neither its predictors nor its ridge penalty depended on population-derived scaling.

Evaluation combined passive and interventional tests. We used an unforced rollout and a pulse excluded from participant-specific calibration. Its amplitude was $2.25$, beyond the calibration range of $[-1.70,1.70]$ but within the perturbational-pretraining range of $[-2.30,2.30]$, and its duration was 120 simulation steps. The primary pulse test therefore assesses transfer beyond participant-specific calibration, not extrapolation beyond population-pretraining support. A supplementary test used an otherwise identical pulse of amplitude $3.00$, just outside population-pretraining support.

Structural quantities were evaluated on a common state-input grid. Finite-run empirical occupancy and response distributions were estimated from 40 stochastic replicates using 80 bins over $[-2.3,2.3]$ and a $10^{-8}$ pseudocount. Occupancy was calculated after a 1,000-step burn-in from 5,000-step unforced simulations initialised in the left basin. Response-distribution Jensen-Shannon divergence was evaluated every 20 steps across the complete trajectory and averaged over those time points. It therefore compares marginal state distributions at individual times, not the joint law over whole trajectories.

Dose-transition curves used 12 pulse amplitudes from $0.5$ to $2.8$, with 40 stochastic replicates per amplitude. Their upper range therefore extended beyond the perturbational-pretraining limit of $2.3$, whereas the primary test pulse of $2.25$ remained within population-pretraining support. The same known diffusion level was used for the true and learned systems. Distributional differences therefore reflected reconstructed-drift error rather than diffusion mismatch, apart from finite-sample Monte Carlo variation.

Transition outcomes are classified against the saddle of the true potential for the
single held-out pulse, the primary stochastic dose-transition curve, and the robustness
sweep. This common physical threshold asks whether every model predicts the switch that
actually occurs in the generating system. In each primary analysis a trajectory is counted
as having switched when the mean of its final 61 steps lies on the opposite side of the
threshold from its starting side; the robustness sweep uses only the final sample. As
secondary diagnostics, the output table also reports two model-relative quantities, in
which each model is classified against the saddle of its own recovered potential: the
stochastic dose-transition RMSE, and a deterministic amplitude-curve error computed from
noise-free rollouts. Both ask whether a trajectory switches within the model's internal
landscape rather than in the generating system, and both are defined only when the
recovered landscape contains two minima and an intervening saddle, which does not always
hold at the smallest calibration sizes (Supplementary Table~S4).
Neither appears in the main learning curves. Reporting the two conventions separately keeps physical outcome
prediction distinct from internal landscape consistency.

\subsection{Coverage and excitation}

\begin{figure}[t]
\centering
\includegraphics[width=\textwidth]{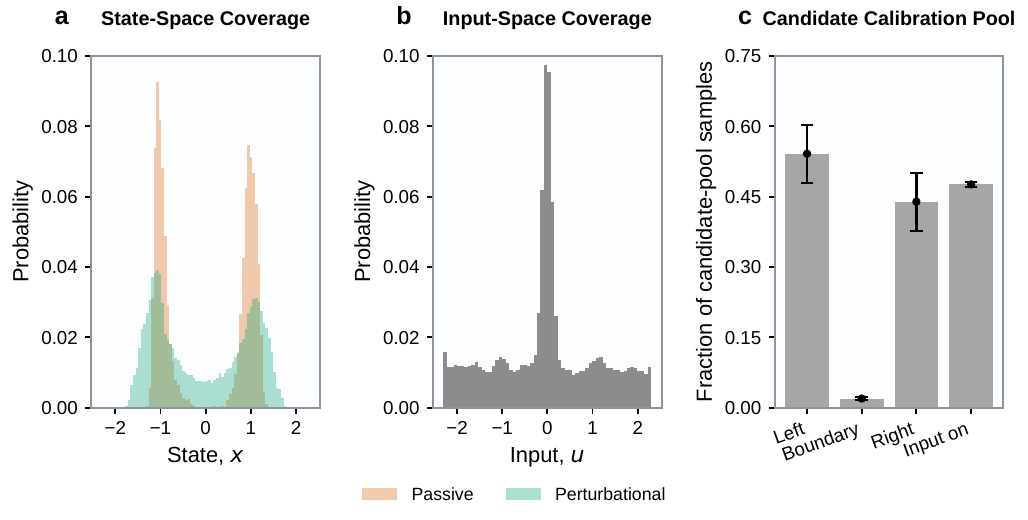}
\caption{\textbf{Coverage and excitation diagnostics.}
\textbf{a}, Empirical state occupancy during population pretraining. Passive trajectories (grey) are concentrated around the two attractors, whereas perturbational trajectories (green outline) additionally sample the transition region between the wells.
\textbf{b}, Input distribution during perturbational pretraining, showing excitation over positive and negative amplitudes together with frequent near-zero input.
\textbf{c}, Composition of the candidate calibration pools for held-out systems. Bars show the mean fraction of pool observations in the left basin, near the basin boundary, in the right basin, and during periods with $|u|>0.15$; error bars denote standard errors across held-out systems}
\label{fig:coverage}
\end{figure}

Before comparing transfer performance, we verified that the two pretraining regimes differed in the information relevant to identification (Fig.~\ref{fig:coverage}). Passive trajectories were concentrated around the two stable wells and provided almost no observations of the intervening transition region (Fig.~\ref{fig:coverage}a). Perturbational trajectories continued to sample both wells but also populated states between them. Their inputs spanned positive and negative amplitudes, with frequent near-zero values and broad coverage of the range used to tilt the landscape (Fig.~\ref{fig:coverage}b).

The candidate calibration pool for each held-out system comprised two input-driven episodes initialised in opposite basins. Across the full pool, states covered the left basin, the right basin, and periods with $|u|>0.15$, but few observations fell near the narrow basin boundary (Fig.~\ref{fig:coverage}c). These diagnostics describe the pool from which each calibration set was sampled, not the composition of the independently selected sets themselves. The candidate pool therefore did not densely reconstruct the transition region; successful few-shot recovery from its sparse subsets had to rely on population-level structure acquired during pretraining.

Three properties of the metrics bound what they can show. The attractor-location error
compares the two deepest minima of the true and recovered potentials and adds a penalty
when fewer than two are found, so it detects a collapsed landscape but is blind to a
spurious third attractor. Passive-rollout RMSE compares a deterministic model rollout with
one stochastic realisation of the truth, so its absolute value has an irreducible
process-noise floor and, close to the transition threshold, a single fortunate basin switch
can dominate a participant's score; because all three models are scored against the same
reference trajectory, only the comparison between models is interpretable. Finally, the
simulated state is clipped at $|x|\leq2.3$, so the ground truth is a clipped
Euler-Maruyama chain rather than the unconstrained diffusion. The clip was never active in
any reported condition. The largest state visited in the population data was $1.84$, and no
trajectory reached the boundary under the amplitude-$2.25$ or amplitude-$3.00$ pulses or
anywhere on the dose-response grid. The same holds for the fitted models: across all three
model classes and every calibration size, the largest state reached in any rollout or
stochastic ensemble was $2.04$, attained by the from-scratch model at two calibration
evaluations, where its fit is poorest.

These diagnostics connect the simulation to the identifiability argument in Section~\ref{sec:framework}. Passive data provide high density where the system naturally resides but no excitation of the input direction needed to identify $b_s$ and the controlled vector field (see Eq.~\ref{eq:double-well}). Perturbational pretraining increases both state-space coverage and input-space excitation, narrowing the set of controlled dynamics compatible with the observations.

The primary comparison changes state-space coverage and input excitation simultaneously. We therefore used two matched supplementary ablations to separate their contributions. The state-coverage model was trained on states visited during driven trajectories but received zero inputs and the corresponding unforced drift. The input-excitation model received varied inputs and their analytical controlled drift at states sampled from passive trajectories. Input excitation improved identification more than transition-state coverage alone on the input-dependent criteria, most clearly for controlled-flow RMSE, which fell from $3.12\pm1.22$ to $1.61\pm0.09$ at two calibration evaluations and was lower in 5 of 5 runs. Unforced landscape geometry showed no such advantage in the few-shot regime, as expected for a quantity determined by the unforced field. Combining excitation with driven state coverage produced the strongest population-pretrained model on every metric and at every calibration size (Supplementary Fig.~S1a,b and Table~S3).

\subsection{Mechanistic validation in a held-out system}

\begin{figure*}[!t]
\centering
\includegraphics[width=\textwidth]{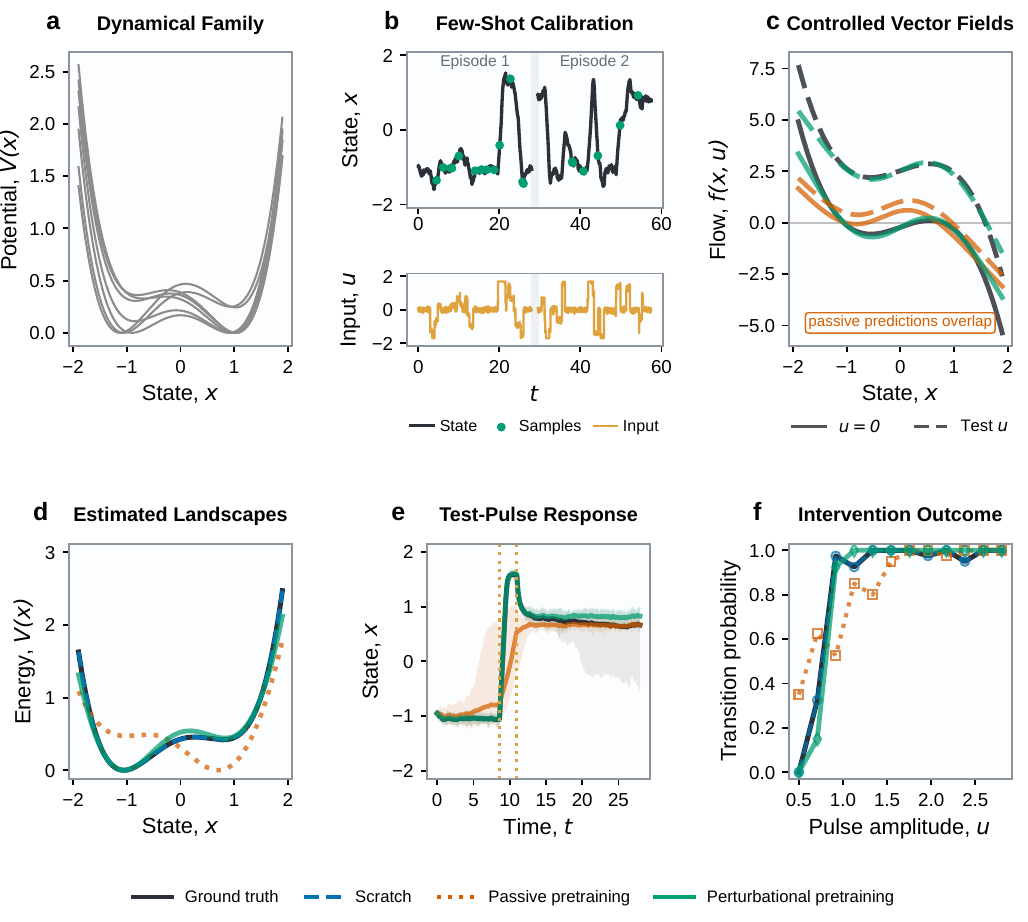}
\caption{\textbf{Mechanistic validation of perturbational transfer in a held-out system.}
\textbf{a}, Representative potential functions from the population family.
\textbf{b}, Two calibration episodes for one held-out system, initialised in opposite basins, with 20 selected state-input locations and their analytical drift targets across the episodes; dark grey denotes the state trajectories, amber the input drives, green markers the locations used to estimate the participant embedding, and the pale vertical gap separates the episodes.
\textbf{c}, True and reconstructed controlled vector fields for zero input (solid) and a pulse amplitude excluded from participant-specific calibration (dashed). Perturbational pretraining reproduces the input-dependent displacement, whereas the passive model's controlled and unforced predictions largely overlap.
\textbf{d}, Ground-truth and reconstructed unforced energy landscapes. Perturbational pretraining and the correctly specified from-scratch model recover the minima and central barrier, whereas passive pretraining produces a distorted landscape.
\textbf{e}, Ensemble response to the pulse excluded from participant-specific calibration. Lines show ensemble means and shaded regions show 90\% intervals; vertical dotted lines mark pulse onset and offset. Perturbational pretraining closely reproduces the timing and distribution of the transition, whereas passive pretraining produces a slower and differently distributed response.
\textbf{f}, Transition probability as a function of pulse amplitude. Perturbational pretraining and the from-scratch model recover the sharp intervention threshold, whereas passive pretraining predicts excessive transition probability below the true threshold and an overly gradual dose-transition relationship}
\label{fig:mechanistic-validation}
\end{figure*}

Figure~\ref{fig:mechanistic-validation} follows one held-out system calibrated with 20 oracle drift evaluations. The population contains related but non-identical double-well landscapes (Fig.~\ref{fig:mechanistic-validation}a), so transfer requires locating the new system within this family rather than copying a single population average. The two calibration episodes contain state trajectories and time-varying inputs; only the highlighted state-input locations and their analytical drift targets were used to estimate the participant embedding (Fig.~\ref{fig:mechanistic-validation}b).

The key structural comparison is the controlled vector field (Fig.~\ref{fig:mechanistic-validation}c). Under $u=0$, both pretrained models capture parts of the unforced flow. Under the pulse amplitude excluded from participant-specific calibration, however, the passive model's controlled and unforced predictions largely overlap. It therefore fails to reproduce the input-induced displacement of the vector field. The perturbationally pretrained model provides a substantially closer approximation under both input conditions. Its advantage is therefore not merely a better fit to one trajectory, but improved recovery of the state-input law $f(x,u)$.

The unforced energy landscapes provide the corresponding geometric view (Fig.~\ref{fig:mechanistic-validation}d). Perturbational pretraining recovers the two minima, the central barrier, and the overall landscape shape, whereas passive pretraining produces a markedly distorted landscape in regions weakly constrained by passive data. At this calibration size, the correctly specified from-scratch model also recovers the unforced landscape closely. Because $f(x,0)=-dV/dx$, errors in the estimated drift accumulate into errors in basin geometry and barrier height.

The response to the held-out pulse shows why a binary transition outcome is insufficient (Fig.~\ref{fig:mechanistic-validation}e). The true system transitions from the left to the right basin, and the perturbationally pretrained model reproduces the timing of that transition almost exactly. It does not reproduce the spread: its ensemble is markedly narrower than the true one and settles slightly higher, so it recovers when the transition happens more accurately than how variable the outcome is. The passive model also leaves the initial basin in this example, but crosses later and with a different final distribution. Moreover, because its controlled vector field changes little with input, this transition does not reflect accurate recovery of the input-dependent law.

The dose-transition curve makes this error more apparent (Fig.~\ref{fig:mechanistic-validation}f). Perturbational pretraining approximates the sharp increase in transition probability as pulse amplitude crosses the effective threshold. In contrast, passive pretraining assigns substantial transition probability to amplitudes below the true threshold and produces an overly gradual dose-transition relationship. The correctly specified from-scratch estimate is competitive because 20 oracle drift evaluations are sufficient to recover the cubic input-response law.

Together, the panels connect the formal framework to a prospective transfer test. Panel b shows the limited participant-specific calibration; panels c and d assess dynamical structure; panel e assesses the response distribution; and panel f assesses the intervention outcome. The central requirement is that the transferred model preserve distinctions among inputs according to their consequences, rather than merely reproduce frequently observed states or a single binary transition outcome.

\subsection{Few-shot transfer across complementary criteria}

\begin{figure*}[!t]
\centering
\includegraphics[width=\textwidth]{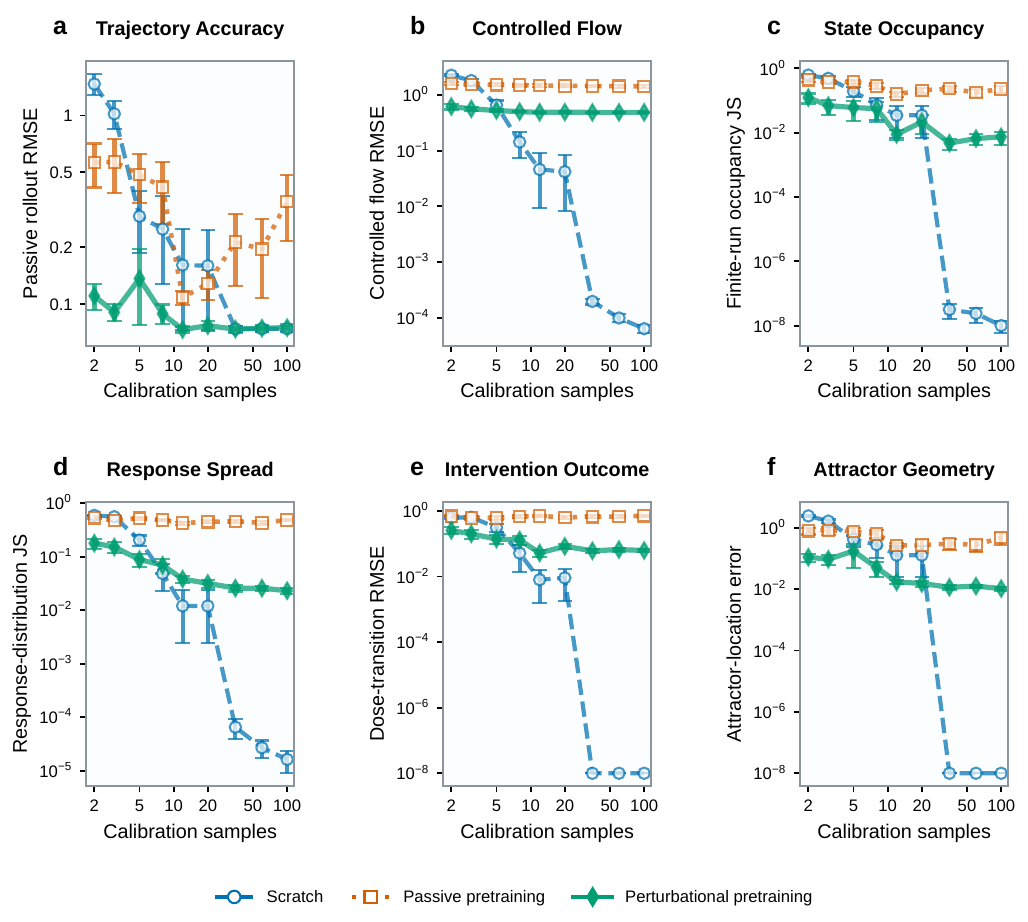}
\caption{\textbf{Few-shot transfer across complementary validation criteria.}
Performance is shown as a function of independently sampled calibration-set size for a regularised cubic model fitted from scratch (blue), a shared model pretrained on passive trajectories (orange), and a shared model pretrained on perturbational trajectories (green). Each calibration sample includes an analytical drift target. Points show means across 20 held-out systems and error bars show standard errors; both axes are logarithmically spaced.
\textbf{a}, RMSE of a deterministic passive rollout against one stochastic reference trajectory.
\textbf{b}, RMSE of the controlled vector field over a common state-input grid.
\textbf{c}, Jensen-Shannon divergence between true and model-generated empirical occupancy distributions under the specified finite simulation protocol. This divergence is bounded above by $\ln 2$ and remains finite where a model assigns no mass, so unlike a Kullback-Leibler divergence its magnitude does not depend on the histogram smoothing constant.
\textbf{d}, Jensen-Shannon divergence between response distributions under a pulse excluded from participant-specific calibration.
\textbf{e}, RMSE between true and predicted dose-transition curves.
\textbf{f}, Error in recovered attractor locations.
Perturbational pretraining provides its clearest advantage with two to five oracle drift evaluations, whereas passive pretraining remains poor on input-dependent criteria and the correctly specified from-scratch model becomes competitive with larger calibration sets}
\label{fig:few-shot-metrics}
\end{figure*}

Figure~\ref{fig:few-shot-metrics} extends the comparison to all 20 held-out systems and all calibration sizes. Sets at different sizes were sampled independently, so the curves compare performance across sample counts rather than tracking nested observations. The apparent ranking of the models depends strongly on the evaluation criterion.

For deterministic passive-rollout RMSE against a stochastic reference trajectory, perturbational pretraining yields low error with only two to three oracle drift evaluations and remains stable as calibration grows (Fig.~\ref{fig:few-shot-metrics}a). The from-scratch model improves rapidly and approaches the perturbational model with larger sets, whereas passive pretraining remains less accurate. This demonstrates an advantage in the extreme few-shot regime, but the rollout metric alone neither identifies its source nor avoids the process-noise floor.

The structural metrics identify the source and limits of this advantage. Controlled-flow RMSE is lower for perturbational than for passive pretraining across all calibration sizes because only the perturbational model learned during population pretraining how $u$ changes the drift (Fig.~\ref{fig:few-shot-metrics}b). The correctly specified from-scratch model becomes more accurate by eight calibration evaluations and approaches zero error with larger sets.

Finite-run empirical-occupancy Jensen-Shannon divergence shows a related few-shot pattern (Fig.~\ref{fig:few-shot-metrics}c): perturbational pretraining provides a useful approximation with two to five evaluations, whereas the scratch model becomes competitive as calibration increases. These occupancy estimates are conditional on the stated burn-in, simulation length, and left-basin initialisation; stationarity or convergence to an invariant distribution was not established.

The intervention-specific metrics likewise show the clearest separation at the smallest calibration sizes. Perturbational pretraining achieves lower response-distribution Jensen-Shannon divergence with two to five oracle drift evaluations (Fig.~\ref{fig:few-shot-metrics}d), and dose-transition RMSE follows the same pattern (Fig.~\ref{fig:few-shot-metrics}e). By eight evaluations, the correctly specified scratch model is already competitive on both criteria. Its few-shot weakness is structural rather than merely quantitative: with two or three evaluations it recovered a bistable landscape in none of the 20 held-out systems, whereas perturbational pretraining recovered one in 18 (Supplementary Table~S4). Attractor-location error further clarifies the trade-off (Fig.~\ref{fig:few-shot-metrics}f): perturbational pretraining provides useful geometric accuracy with little calibration, whereas the scratch model becomes extremely precise once sufficient participant-specific information is available.

Supplementary analyses tested whether these findings depended on the primary random seed, frozen shared weights, or evaluation only within the population-pretraining input range. Across five independent end-to-end runs, the few-shot advantage of perturbational pretraining was reproduced for controlled-flow and response-distribution errors (Supplementary Fig.~S1a,b). Allowing all weights of the passive model to adapt improved its performance but did not remove the few-shot gap. For the amplitude-$3.00$ pulse just outside population-pretraining support, the perturbational model also retained lower response-distribution and transition-probability errors than the passive variants (Supplementary Fig.~S1c,d). This result supports local extrapolation beyond the trained input boundary, but not unrestricted out-of-range generalization.

Taken together, the six primary metrics and supplementary controls show the dissociation predicted by the framework under oracle-drift calibration. A model can appear plausible near familiar trajectories or reproduce an individual transition while failing to approximate controlled flow, response distributions, or intervention thresholds. Perturbational pretraining transfers input-dependent population structure and is most beneficial when only two to five participant-specific drift evaluations are available; its advantage is not a superior asymptotic optimum.

%% file: sections/04_discussion.tex
\section{Discussion}
\label{sec:discussion}

Our proof-of-principle simulation yields one central result: population-level exposure to controlled inputs enabled few-shot transfer of input-dependent dynamics. With two to five participant-specific oracle drift evaluations, perturbational pretraining provided substantially better recovery of the controlled vector field, effective energy landscape, stochastic pulse response, and dose-transition relationship than passive pretraining. The correctly specified model fitted from scratch became competitive as calibration increased, so the advantage was not asymptotic. In the few-shot regime, however, the perturbational model's advantage held: we reproduced it across five independent end-to-end runs, it remained after full fine-tuning of the passive network, and matched ablations clarified its source. With two participant-specific evaluations, input excitation alone reduced controlled-flow RMSE relative to transition-state coverage alone from $3.12\pm1.22$ to $1.61\pm0.09$ (mean $\pm$ SD across five independent runs), a factor of $1.94$, with the lower value in 5 of 5 runs; absolute transition-probability error fell from $0.67\pm0.14$ to $0.48\pm0.09$, also in 5 of 5 runs, and response-distribution divergence from $0.56\pm0.06$ to $0.48\pm0.05$ in 4 of 5 runs. The two information sources were not interchangeable across all criteria. Unforced landscape RMSE did not separate them in the few-shot regime ($1.48\pm0.89$ for coverage versus $1.22\pm0.36$ for excitation, with excitation lower in only 2 of 5 runs), as expected for a quantity determined by the unforced field rather than by input coupling. On that metric, excitation separated consistently from coverage only at 20 evaluations, where it was lower in 5 of 5 runs. Combining excitation with driven state coverage gave the lowest error on all six metrics in 5 of 5 runs and at every calibration size. The perturbational model also retained its advantage for one input amplitude just outside population-pretraining support. Supplementary Fig.~S1 and Table~S3 give the full comparison.

This result follows from the identifiability structure of our simulation. Participant-specific input coupling enters only through terms multiplied by $u$; under the passive condition $u=0$, those parameters disappear from the observed dynamics. Passive data can constrain the unforced field over visited states, but cannot identify input gain without additional assumptions or informative input variation. They are not, however, wholly uninformative about responses to input: variability in stimulation-response strength and prediction accuracy across brain regions has been explained by a functional connectivity measure computed at rest \citep{yang2021modelling}. Such a measure is an instance of the additional assumptions just noted; it constrains relative response magnitude across regions rather than identifying an individual's input coupling. Controlled inputs provide the required excitation, although identification still depends on their amplitude, timing, and state-space coverage \citep{Ljung1999-bv,Villaverde2019-qf,schoukens2019nonlinear}. Perturbational pretraining is useful here because it transfers population-level regularities in how inputs modify the dynamics, reducing what must be learned separately for a new participant.

Our findings also motivate evaluating perturbational validity at three complementary levels. Time-series metrics assess forecasts over a defined horizon; structural metrics assess the controlled field, stable states, occupancy, and transition geometry; and perturbational metrics assess response distributions and dose dependence under held-out inputs. None is sufficient alone. In our simulation, the passive model occasionally predicted the correct binary transition despite having an incorrect controlled field, response timescale, and dose-transition curve. Conversely, passive-rollout RMSE compared a deterministic rollout with one stochastic realization and therefore contained an irreducible process-noise component. More generally, short-horizon accuracy can obscure errors in attractor geometry and invariant statistics \citep{Hemmer2025-xa}. We therefore recommend that intervention-relevant validation establish agreement across the three layers rather than rely on a single forecast or endpoint.

This proposal complements observational evaluation of brain foundation models. REVE, BrainLM, and Brain-JEPA demonstrate transfer through masked reconstruction, forecasting, or downstream prediction \citep{elouahidi2025reve,Ortega-Caro2024-uy,dong2024brainjepa}. Prospective precedents also exist: population-scale neural models have predicted responses in new animals and stimulus domains, closed-loop models have selected stimuli for subsequent experimental testing, and dynamic input-output models have predicted responses to electrical stimulation \citep{wang2025foundation,walker2019inception,yang2021modelling}. Perturbation-prediction work in other biological domains further emphasises comparison with simple, well-specified baselines \citep{ahlmann2025deep}. The contribution here is therefore not the claim that neural models have never predicted perturbation responses, but the proposal that preservation of the controlled law should be an explicit transfer target evaluated at trajectory, structural, and perturbational levels.

This distinction is particularly relevant to dynamical accounts of mental health, where alternative stable states, resilience, and transitions have been proposed as organising concepts for psychiatric and psychotherapeutic change \citep{Scheffer2024-mq,Gelo2016-dx,klocek2024applying}. Network control theory provides a formal language for asking how inputs may move neural systems towards target states \citep{Gu2015-qx,Tang2018-uc,Muldoon2016-hq}. In mental health research, interventions have been formalised as network-control problems; brain controllability has been related to MDD and to genetic and familial risk; and individualised control properties have been used to predict electroconvulsive-therapy response \citep{Stocker2023-rz,Hahn2023-ls,Hahn2023-tg}. Such analyses motivate individualised control models but do not substitute for prospective validation, because controllability estimates depend on the assumed dynamics, connectivity representation, and control objective \citep{tu2018warnings,manjunatha2024controlling}. Perturbational validity asks the corresponding empirical question: does the individualised model predict a response to an input that was not used for its calibration?

Empirically, a perturbation need not be direct neurostimulation. Timed sensory events, cognitive tasks, pharmacological challenges, behavioural manipulations, psychotherapy components, and stimulation protocols can all provide informative state-input variation when their timing and dose are defined. This view is consistent with adaptive N-of-1 designs in which selected perturbations calibrate individualised surrogate models and subsequent observations test their counterfactual predictions \citep{kheirkhah2026reengineering,Hofmann2025-mm}. We therefore propose that prospective benchmarks be integrated into the evaluation of brain foundation models intended for intervention: a population model is adapted using limited data from a new participant and then tested on perturbation types, amplitudes, timings, spatial patterns, or initial states excluded from participant calibration. By including tests outside population-pretraining support and reporting calibrated predictive uncertainty, such benchmarks would provide stronger evidence.

Several limitations bound our current findings. Our simulation used a one-dimensional, stationary bistable gradient system with directly observed state, a scalar input, known diffusion, and a narrowly defined synthetic population. We therefore did not test the observation model of the general framework. Further, we pretrained and calibrated on analytical drift targets, isolating transfer of the controlled law from the harder problem of estimating it from noisy, discretely sampled measurements, and we estimated occupancy from finite simulations initialized in one basin, without demonstrating convergence to an invariant distribution. We chose these simplifications to obtain a controlled proof of principle, but future work should test whether the few-shot advantage survives drift estimation from noisy, partially observed recordings. On the testing side, we used an out-of-support test at only one nearby amplitude. Here, stronger positive pulses approach the imposed state boundary and would not provide a clean graded extrapolation test. Our internal comparison holds across repeated runs, full passive fine-tuning, and matched coverage-excitation ablations, but our scratch baseline was correctly specified for the generating cubic family, whereas empirical models will face model and architecture mismatch. Future research should therefore include graded extrapolation tests and baselines that are not correctly specified for the generating system. Real neural systems are also high-dimensional, partially observed, non-stationary, and often non-gradient \citep{Durstewitz2023-ta}. In such settings, a unique scalar landscape may not exist; controlled vector fields, generators, committor functions, transition operators, or quasipotentials may be more defensible targets \citep{Freidlin2012-wc,nolting2016balls,E2010-xz}. We therefore interpret the numerical advantage in our study as evidence for an evaluation principle, not as an estimate of the benefit expected in neuroimaging or clinical data.

Taken together, our findings support a specific conclusion. When intervention-relevant parameters vanish under the passive condition, passive observations alone cannot identify or transfer them. Recovering them requires informative perturbations or equivalent prior assumptions. In our few-shot oracle-drift setting, perturbational pretraining supplied that information and improved recovery of controlled dynamics. Brain foundation models should therefore be described as perturbationally valid only when they predict the consequences of actively driving the system, evaluated under held-out inputs and across all three validation layers.

%% file: sections/05_conclusion.tex
\section{Conclusion}
\label{sec:conclusion}

Brain foundation models derive their value from transferring population-level structure to new individuals with limited data. For experimental or clinical intervention, however, observational accuracy is necessary but insufficient: a model must preserve the controlled dynamics that determine how inputs alter future trajectories. We formalised this requirement as perturbational validity, assessed through time-series accuracy, dynamical-structure similarity, and perturbational-response similarity. In our controlled bistable-system proof of principle, perturbational pretraining enabled few-shot recovery of controlled dynamics and intervention outcomes, an advantage that held across repeated training runs, full passive fine-tuning, matched coverage and excitation ablations, and a local input-extrapolation check. We therefore argue that the relevant benchmark is prospective: models intended for experiment design, stimulation, or treatment selection should be tested on perturbations withheld from participant-specific calibration.

%% file: sections/06_declarations.tex
\backmatter
\section{Statements and Declarations}

\subsection{Funding}
This work was funded in part by the German Research Foundation (DFG) through SFB/TRR 393 (project grant no. 521379614) and through the Excellence Cluster EXC3066 ``The Adaptive Mind'' (project grant no. 533717223), by a research grant from the University Hospital of Gießen and Marburg (UKGM; no. 1/2024 MR), and by ERA-NET NEURON JTC 2024 (BRAWO project, grant no. ER-2024-23684536). The funders had no role in study design, data generation, analysis, interpretation, manuscript preparation, or the decision to submit the work for publication.

\subsection{Competing interests}
The authors have no competing interests to declare that are relevant to the content of this article.

\subsection{Ethics approval}
Not applicable. This study used synthetic data generated by simulation and involved no human participants or animals.

\subsection{Consent to participate}
Not applicable.

\subsection{Consent for publication}
Not applicable.

\subsection{Data availability}
No empirical data were used. All data analysed here are synthetic and are regenerated deterministically by the simulation code from the reported random seeds.

\subsection{Materials and code availability}
All code used to generate the simulations, analyses, and figures is publicly available at \url{https://github.com/JoseAlanis/perturbational_validity_simulation}. No other materials were used.

\subsection{Author contributions}
Conceptualisation, methodology, and validation: HJ, JCGA, and SA. Formal analysis: HJ and JCGA. Writing---original draft: HJ, JCGA, and SA. Writing---review and editing of the final manuscript: JCGA, SA, MR, BS, MK, SGH, TH, and HJ.

\subsection{Use of AI-assisted tools}
The simulation was first implemented in MATLAB; that prototype is included in the code repository. The Python implementation used for the results reported here was developed with the assistance of two coding tools, Claude Code (Anthropic; Claude Opus 5) and OpenAI Codex (v0.151.0; model gpt-5.6-sol), which helped draft parts of the simulation code and its validation tests. The same tools were also used for proofreading the final manuscript text against typos and errors. All code was reviewed, executed, and verified by JCGA and HJ.

%% file: supplementary/supplement.tex
\ifdefined\standalonesupplement\else\section*{Supplementary Material}\fi
\renewcommand{\thetable}{S\arabic{table}}
\renewcommand{\thefigure}{S\arabic{figure}}
\renewcommand{\theHtable}{S\arabic{table}}
\renewcommand{\theHfigure}{S\arabic{figure}}
\setcounter{table}{0}
\setcounter{figure}{0}

\subsection*{Supplementary Methods and Robustness Checks}

Table~\ref{tab:supp-generative} lists the generative, sampling, and evaluation
parameters. All quantities are dimensionless. Each population trajectory
contained 1,500 retained steps after 150 burn-in steps. Held-out calibration
used two episodes of 1,400 steps initialized in opposite basins. Population
pretraining and participant calibration used analytical drift evaluations,
rather than drift estimates obtained from stochastic state increments.

Table~\ref{tab:supp-training} gives the model and optimization specification.
No validation set or data-dependent early stopping was used. Hyperparameters
were fixed before the reported analyses and were not selected separately for
the passive and perturbational pretraining regimes. Gradient updates used Adam
with $\beta_1=0.9$, $\beta_2=0.999$, and $\epsilon=10^{-8}$.

Stochastic metrics compare the true and modelled systems under the same
Gaussian innovations. Sharing the noise removes the sampling luck that would
otherwise contribute to every model-to-model difference: a perfect model then
attains exactly $0$ on the response-distribution and dose-transition metrics,
which we verified by evaluating the true drift against itself ($0.0000$ on
both). Had the two sides been simulated independently, the same check gives
$0.048$ response divergence and $0.027$ dose-transition RMSE with 40
replicates and 80 bins, so differences below roughly that size would have been
indistinguishable from noise. Reported values are therefore attributable to
reconstructed-drift error rather than to independent sampling.

The primary comparison froze the shared network and adapted only a new
participant embedding. To test whether this restriction drove the result, a
stronger passive control updated all network weights, biases, and the new
embedding during calibration. Full adaptation improved the passive model but
did not remove the few-shot gap, indicating that sparse calibration only
partially repaired an input pathway that received no population-level
excitation.

Two matched ablations separated state-space coverage from input excitation.
The \emph{state-coverage} model used states visited during driven
trajectories, but was trained with $u=0$ and the corresponding unforced drift.
It therefore observed the transition region without information about input
coupling. The \emph{input-excitation} model used states from passive
trajectories paired with varied synthetic inputs and their analytical
controlled drift. It therefore received information about input effects near
the attractors without driven transition-state coverage. Both controls used
the same number of samples, architecture, initialization scheme, and optimizer
as the original models.

The complete pipeline was repeated for five independent training seeds, with
20 held-out systems per seed and calibration sizes of 2, 5, and 20. Table~\ref{tab:supp-ablation}
reports the two ablations and their combination at each calibration size. Input
excitation alone outperformed transition-state coverage alone on the
input-dependent criteria at every calibration size, most clearly for
controlled-flow RMSE, where it was lower in 5 of 5 runs at all three sizes and
by a factor of $1.94$ to $1.99$. The ordering was metric-dependent rather than
uniform: unforced landscape RMSE did not separate the two ablations in the
extreme few-shot regime and favoured excitation only at 20 evaluations. The
combination of excitation and driven state coverage gave the lowest error on
every metric, at every calibration size, in 5 of 5 runs. The correctly
specified model fitted from scratch became competitive as participant-specific
calibration increased.

The model-relative transition diagnostics require the recovered potential to have
two minima with an intervening saddle. Table~\ref{tab:supp-collapse} reports how
often that condition failed. With two or three calibration evaluations the
from-scratch cubic model recovered a single well in every held-out system. Its
functional form is correct, but two or three drift evaluations do not determine
five coefficients, so the fit collapses to one basin. Perturbational pretraining
recovered a double well in 18 of 20 systems at the same budget and in all 20 from
twelve evaluations onward. Passive pretraining failed in roughly half the systems
in the few-shot regime and still failed in 3 of 20 at 100 evaluations, consistent
with passive data leaving the transition region weakly determined however much
calibration is supplied. Because the primary dose-transition metric classifies
every model against the saddle of the true potential, it stays defined for these
collapsed reconstructions, whereas the model-relative diagnostic does not.

The amplitude-2.25 test pulse was excluded from participant calibration but
remained within the population-pretraining range. We therefore also evaluated
an otherwise identical amplitude-3.00 pulse just beyond the population-
pretraining range. The perturbational model retained lower response-
distribution and transition-probability errors than the passive variants at
this first out-of-support amplitude. This result supports only local
extrapolation beyond the trained input boundary. Stronger positive pulses
rapidly drove the clipped bistable system toward its state boundary and were
therefore less informative as a graded stress test.

\makeatletter
\setlength{\@fptop}{0pt}
\setlength{\@fpbot}{0pt plus 1fil}
\makeatother

\clearpage

\begin{figure*}[!t]
\centering
\includegraphics[width=\textwidth]{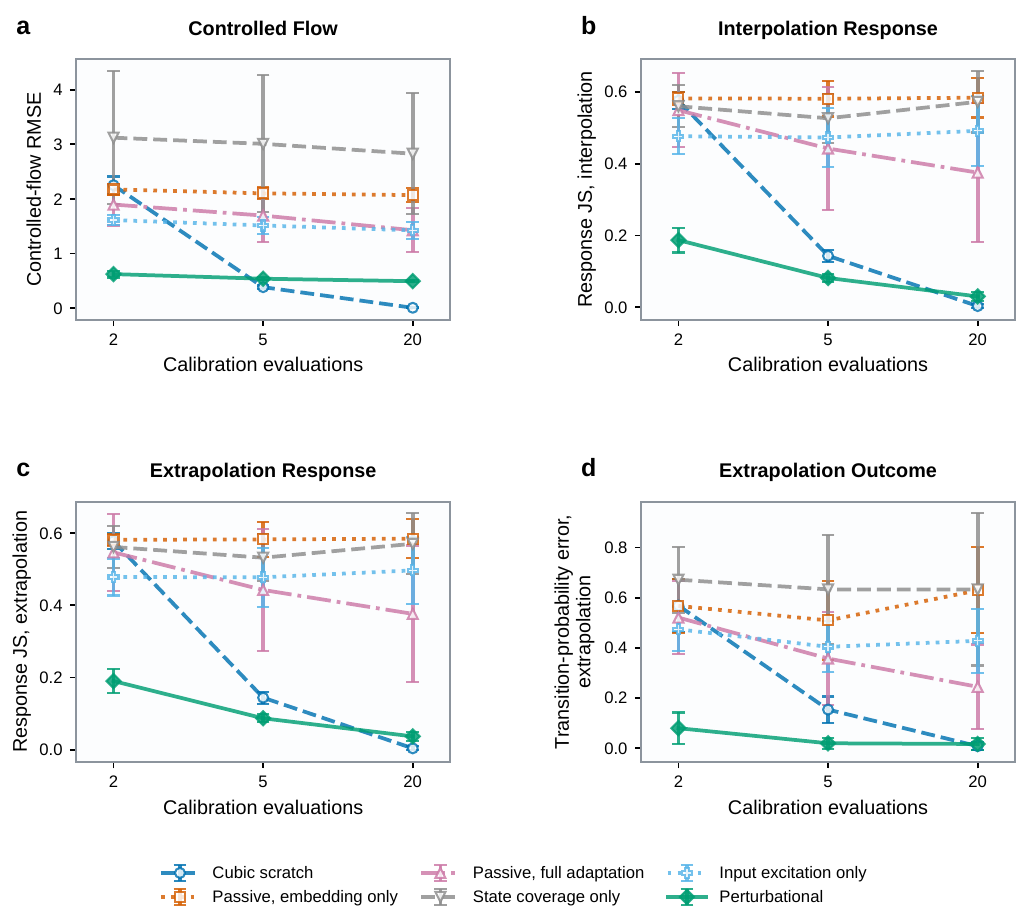}
\caption{\textbf{Repeated-run robustness, adaptation controls, identification ablations, and local input extrapolation.}
For each run, metrics were first averaged across 20 held-out systems; points
show the means and error bars the standard deviations of these run-level means
across five independent end-to-end runs.
\textbf{a}, Controlled-flow RMSE.
\textbf{b}, Response-distribution Jensen-Shannon divergence for the amplitude-
2.25 interpolation pulse.
\textbf{c}, Response-distribution Jensen-Shannon divergence for the amplitude-
3.00 pulse just outside population-pretraining support.
\textbf{d}, Absolute transition-probability error for the same out-of-support
pulse}
\label{fig:supp-controls}
\end{figure*}

\clearpage

\begin{sidewaystable}[p]
\caption{\textbf{Generative, sampling, and evaluation parameters.}}
\label{tab:supp-generative}
\centering
\small
\begin{tabularx}{\textheight}{L{0.95}L{0.85}L{1.20}}
\toprule
\textbf{Quantity} & \textbf{Value} & \textbf{Role} \\
\midrule
Training and held-out systems & 60 and 20 & Population pretraining and participant-level evaluation \\
Integration step and state clip & $dt=0.02$, $|x|\leq2.3$ & Euler-Maruyama integration \\
Participant parameters & $a=1$; $c\in[0.75,1.30]$; $d\in[-0.28,0.28]$; $b\in[0.60,1.45]$; $\sigma=0.14$ & Uniform population family \\
Training trajectory & 1,500 retained steps; 150 burn-in steps & Equal duration in all pretraining regimes \\
Perturbational pretraining & pulse probability 0.035; duration 35-90 steps; amplitude $[-2.30,2.30]$; input-noise SD 0.10 & Population-level controlled excitation \\
Calibration episodes & 2 episodes of 1,400 steps & Opposite initial basins \\
Calibration input & 18 pulses per episode; duration 30-75 steps; amplitude $[-1.70,1.70]$; input-noise SD 0.08 & Participant-specific identification pool \\
Calibration sizes & 2, 3, 5, 8, 12, 20, 35, 60, 100 & Independently sampled sets \\
Perturbed calibration fraction & 0.50 where available & Samples with $|u|>0.15$ \\
Interpolation pulse & amplitude 2.25; duration 120 steps; onset at step 430 & Outside participant-calibration support but inside population-pretraining support \\
Extrapolation pulse & amplitude 3.00; duration 120 steps; onset at step 430 & Outside population-pretraining support \\
State grid & 450 points over $[-1.9,1.9]$ & Potential and flow evaluation \\
Flow-input grid & $-2,-1,0,1,2,2.25$ & Controlled-flow evaluation \\
Occupancy grid & 80 bins over $[-2.3,2.3]$; pseudocount $10^{-8}$ & Distributional metrics \\
Stochastic evaluation & 40 replicates & Response and transition probabilities \\
Finite-run occupancy & 5,000 steps; 1,000-step burn-in & Finite empirical occupancy estimate \\
\bottomrule
\end{tabularx}
\end{sidewaystable}

\clearpage

\begin{table*}[!t]
\caption{\textbf{Model, optimization, and regularization specification.}}
\label{tab:supp-training}
\centering
\small
\begin{tabularx}{\textwidth}{L{0.95}L{0.80}L{1.25}}
\toprule
\textbf{Quantity} & \textbf{Value} & \textbf{Procedure} \\
\midrule
Shared network & two tanh layers, 24 units each & Inputs are standardized state, input, and participant embedding \\
Participant embedding & 3 dimensions & One embedding column per training system \\
Initialization & weights SD 0.10; embeddings SD 0.05; biases zero & Passive and perturbational models share the same initial parameters within a run \\
Pretraining & 55 epochs; batch size 2,048; learning rate $2\times10^{-3}$ & Samples reshuffled each epoch \\
Embedding adaptation & 120 epochs; learning rate $3\times10^{-2}$ & Shared weights frozen; new embedding initialized at zero \\
Full-network passive control & 120 epochs; learning rate $3\times10^{-3}$ & All weights, biases, and the new embedding updated \\
Weight decay & $10^{-5}$ & Applied to network weights \\
Embedding penalty & $10^{-3}$ & Quadratic penalty on embeddings \\
Gradient clipping & global norm 5.0 & Applied before each Adam update \\
Scratch baseline & cubic ridge regression in physical coordinates; penalty $10^{-4}$ & Intercept unpenalized; correctly specified polynomial and linear-input form; no population-derived scaling \\
Primary random seed & 11 & Original single-run analysis \\
Robustness seeds & 11, 23, 37, 51, 71 & Independent parameters, samples, initialization, training, and evaluation \\
\bottomrule
\end{tabularx}
\end{table*}

\clearpage

\begin{table*}[!t]
\caption{\textbf{Matched identification ablations across five independent runs.}
Entries are means and standard deviations of run-level means, where each
run-level mean is first averaged across 20 held-out systems. \emph{Coverage}
used driven states with $u=0$ targets, \emph{excitation} used passive states
with varied synthetic inputs, and \emph{both} is the perturbational model that
receives driven states and their controlled drift. The final column counts the
runs in which excitation gave a lower error than coverage. Values in
parentheses are pulse amplitudes.}
\label{tab:supp-ablation}
\centering
\small
\setlength{\tabcolsep}{5pt}
\begin{tabularx}{\textwidth}{L{1.00}ccccc}
\toprule
\textbf{Metric} & \textbf{$n_{\mathrm{cal}}$} & \textbf{Coverage} & \textbf{Excitation} & \textbf{Both} & \textbf{Runs} \\
\midrule
Controlled-flow RMSE & 2 & $3.12 \pm 1.22$ & $1.61 \pm 0.09$ & $0.62 \pm 0.06$ & 5/5 \\
 & 5 & $3.01 \pm 1.25$ & $1.51 \pm 0.16$ & $0.54 \pm 0.03$ & 5/5 \\
 & 20 & $2.83 \pm 1.11$ & $1.43 \pm 0.15$ & $0.49 \pm 0.02$ & 5/5 \\
\addlinespace
Landscape RMSE & 2 & $1.48 \pm 0.89$ & $1.22 \pm 0.36$ & $0.29 \pm 0.06$ & 2/5 \\
 & 5 & $1.07 \pm 0.85$ & $0.77 \pm 0.27$ & $0.17 \pm 0.04$ & 3/5 \\
 & 20 & $1.02 \pm 0.70$ & $0.46 \pm 0.10$ & $0.10 \pm 0.01$ & 5/5 \\
\addlinespace
Response divergence (2.25) & 2 & $0.56 \pm 0.06$ & $0.48 \pm 0.05$ & $0.19 \pm 0.03$ & 4/5 \\
 & 5 & $0.53 \pm 0.07$ & $0.47 \pm 0.08$ & $0.08 \pm 0.01$ & 3/5 \\
 & 20 & $0.57 \pm 0.09$ & $0.49 \pm 0.10$ & $0.03 \pm 0.01$ & 4/5 \\
\addlinespace
Transition-prob. error (2.25) & 2 & $0.67 \pm 0.14$ & $0.48 \pm 0.09$ & $0.08 \pm 0.06$ & 5/5 \\
 & 5 & $0.62 \pm 0.23$ & $0.42 \pm 0.11$ & $0.02 \pm 0.02$ & 3/5 \\
 & 20 & $0.62 \pm 0.31$ & $0.43 \pm 0.13$ & $0.02 \pm 0.02$ & 4/5 \\
\bottomrule
\end{tabularx}
\end{table*}

\clearpage

\begin{table*}[!t]
\caption{\textbf{Recovered landscapes that were not bistable.}
Number of the 20 held-out systems whose recovered potential had fewer than two
minima, so that no saddle and no model-relative transition threshold could be
defined. The generating systems are bistable by construction.}
\label{tab:supp-collapse}
\centering
\small
\setlength{\tabcolsep}{6pt}
\begin{tabularx}{\textwidth}{cL{1.00}L{1.00}L{1.00}}
\toprule
\textbf{$n_{\mathrm{cal}}$} & \textbf{From scratch} & \textbf{Passive pretraining} & \textbf{Perturbational pretraining} \\
\midrule
2 & 20 & 9 & 2 \\
3 & 20 & 8 & 2 \\
5 & 6 & 9 & 2 \\
8 & 3 & 7 & 1 \\
12 & 1 & 6 & 0 \\
20 & 1 & 2 & 0 \\
35 & 0 & 3 & 0 \\
60 & 0 & 3 & 0 \\
100 & 0 & 3 & 0 \\
\bottomrule
\end{tabularx}
\end{table*}